# Towards synthetic neural networks: Can artificial electrochemical neurons be coupled with artificial memristive synapses?

Ewelina Wlaźlak,[1*] Dawid Przyczyna,[1,2] Rafael Gutierrez,[3*] Gianaurelio Cuniberti,[3,4,5] and Konrad Szaciłowski[1*]

[1]*Academic Centre for Materials and Technology, AGH University of Science and Technology, al. Mickiewicza 30, 30-059 Kraków, Poland*

[2]*Faculty of Physics and Applied Computer Science, AGH University of Science and Technology, al. Mickiewicza 30, 30-059 Kraków, Poland*

[3]*Institute for Materials Science and Max Bergmann Center of Biomaterials, Dresden University of Technology, 01062 Dresden, Germany*

[4]*Dresden Center for Computational Materials Science (DCMS), TU Dresden, 01062 Dresden, Germany*

[5]*Center for Advancing Electronics Dresden, TU Dresden, 01062 Dresden, Germany*

*Corresponding authors e-mails: ewlazlak@agh.edu.pl, rafael.gutierrez@tu-dresden.de, szacilow@agh.edu.pl

# Abstract

The enormous amount of data generated nowadays worldwide is increasingly triggering the search for unconventional and more efficient ways of processing and classifying information, eventually able to transcend the conventional von-Neumann-Turing computational central dogma. It is, therefore, greatly appealing to draw inspiration from less conventional but computationally more powerful systems such as the neural architecture of the human brain. This neuromorphic route has the potential to become one of the most influential and long-lasting paradigms in the field of unconventional computing. The material-based workhorse for current hardware platforms is largely based on standard CMOS technologies, intrinsically following the above mentioned von-Neumann-Turing prescription; we do know, however, that the brain hardware operates in a massively parallel way through a densely interconnected physical network of neurons. This requires challenging the intrinsic definition of the single units and the architecture of computing machines. Memristive and the recently proposed memfractive systems have been shown to display basic features of neural systems such as synaptic-like plasticity and memory features, so that they may offer a diverse playground to implement synaptic connections. Their combination with reservoir computing approaches can further increase their versatility since reservoir networks do not require extra optimization of internal connections. In this review, we address various material-based strategies of implementing unconventional computing hardware: (i) electrochemical oscillators based on liquid metals and (ii) mem-devices exploiting Schottky barrier modulation in polycrystalline and disordered structures made of oxide or perovskite-type semiconductors. Both items (i) and (ii) build the two pillars of neuromimetic computing devices, which we will denote as *synthetic neural networks*. We complement the more experimental aspects of the review with an overview of few atomistic and phenomenological modelling approaches of memdevices as well as of reservoir computing networks. We expect that the current review will be of great interest for scientists aiming at bridging unconventional computing strategies with specific materials-based platforms.

# Table of contents

# 1. Introduction

In recent decades, we are witnessing an explosion of the amount of data accumulated in all spheres of human activity. The upcoming years may bring a new global problem, sometimes called "the information gap" or "information black hole". It originates from the fact that mankind doubles the amount of stored data every two years, but processing and analysis involve only a small fraction of this collected data. The lack of analysis of this "information black holes" may bring to humanity many unexpected and poorly explained events.[1-2] The "Big Data" problem requires, therefore, effective theoretical and practical disruptive new methodologies of data processing and analysis. Contemporary silicon-based devices, based on the Turing architecture, cannot provide energy-efficient computing power to process this data.[1, 3]

Today's computer technologies, largely based on semiconductor materials, have an extraordinarily successful history. However, despite its unprecedented achievements, current classical computational paradigms encompass only a small subset of all computational possibilities.[4] There are a very broad class of computational substrates and environments, mainly physical systems, whose rich dynamics can be exploited as a computing medium.[5] Molecular electronics, which in principle should transfer classical electronics to the molecular scale and give unparalleled advantages as compared to silicon-based electronics, has not yielded any competitive technology so far; however it has provided a unique insight into chemistry and physics at the single-molecule level.[6-7]

Therefore, alternative computing architectures and computational paradigms have to be explored, as well as different computational systems that can be combined into larger systems, an approach referred to as heterotic computing.[8] One of the most critical challenges for classical computing today originates from the memory bottleneck and the high cost (in terms of time and energy) of constant data transfer from memory to CPU and vice versa, which is commonly termed as the von Neumann bottleneck. On the other hand, we encounter the extreme thermodynamic efficiency of the biological brain computing architecture, where information processing and storage are not physically separated processes. Therefore, the direct integration of information processing units and memory is an approach that would significantly reduce these bottlenecks, rendering computing much faster and more energy-efficient.

Unconventional methods of information processing are tempting, but their potential is not fully recognized yet. A special place among these new approaches is occupied by devices, systems, and ideas drawing inspiration from a highly sophisticated system: the brain. Neuromorphic solutions implemented by software engineers have started to revolutionize our world. Neural networks, machine learning, and artificial intelligence (AI) are in the main focus of the automotive and aviation industries as well as of companies specialized in internet-related services. Whenever a large amount of data is produced and needs to be fast processed, neural energy-efficient deep learning algorithms and artificial intelligence play a crucial role, e.g. in the aviation and automotive industry as autonomous control (Tesla, Porsche), or internet applications as in face and speech recognition (Google Vision, Siri). These systems, however, are based on digital electronics, the neuromorphic computing features are realized software-wise, and they cannot mimic the extreme complexity of neural systems, neither in terms of structure nor complexity and performance. The hardware level is the next reasonable step which several companies: Hewlett-Packard (U.S), IBM (U.S.), Intel Corporation (U.S.), Samsung Group (South Korea) are currently exploring.[9-11]

The current approach involving artificial intelligence and deep learning algorithms, resembling the working of the human brain, is based on traditional implementations on a software level, and sometimes also on CMOS implementations of neural architectures.[12-16] Deep learning relies on artificial neural networks that are typically executed on computers based on the conventional von Neumann architecture, operating mostly sequentially and based on Boolean logic.[17] In contrast, the brain's hardware operates in a massively parallel fashion through a densely interconnected physical network of neurons[18] and in an analog manner with diverse modes of information coding.[19-20]

## 2. Memristor-oscillator coupling: Neuron-like hybrid system

The steps towards real *Artificial Intelligence* should involve stochasticity and randomness, the prerequisites of creativity in natural systems. The stochasticity should be present at various levels of the neuromimetic structure – at the device and system level. It seems, therefore, that neuromorphic analog electronics, despite their noisy character, may turn out to be useful in numerous computing tasks. Among all elements, memristors and other memristive devices seem to be, at least temporarily, a viable computational alternative. Therefore, *in materio* computing, despite its drawbacks and obvious limitations, irrespectively of its implementation (memristive, spintronic, molecular), seems to be the only way to overcome various bottlenecks of conventional computing (Figure 1).[21-26]

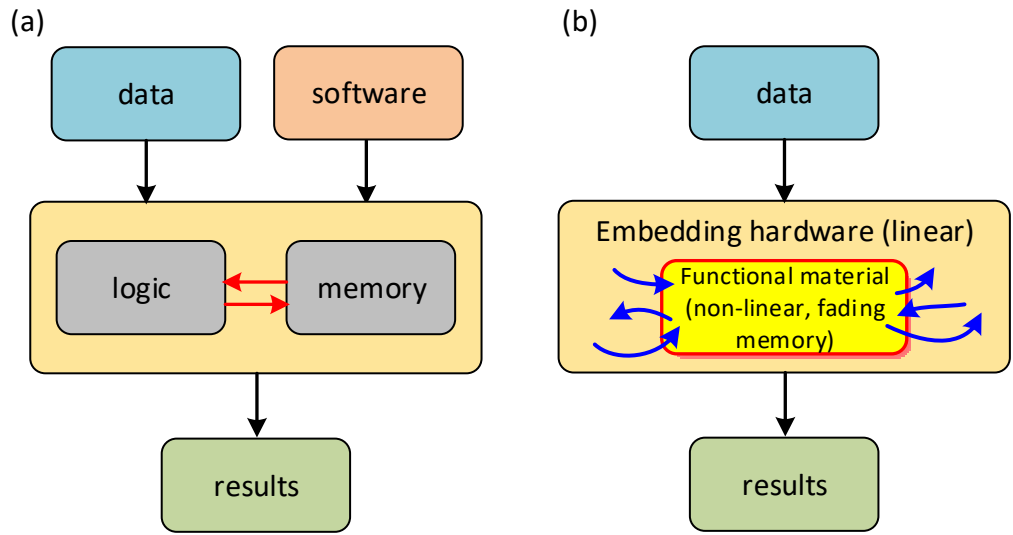

**Figure 1.** Cartoon representation of a classical von Neumann computing architecture (a) versus *in materio* computing system (b). Red arrows indicate the *von Neumann bottleneck* problem, whereas blue arrows show information transfer between the computing material and the interface/readout system.

Memristors and other memristive (or memfractive, *vide infra*) devices bear memory features (volatile and non-volatile), synaptic-like plasticity, and nonlinear electric characteristics. These features are ideally suited for various information processing protocols, including neural networks, reservoir computers, random number generation, and cryptography, including physical unclonable functions. They are also suitable for electric signal transformations due to strongly nonlinear characteristics and the ability to generate higher harmonic frequencies. Their main drawback lies in their passive character – they can sink current and dissipate energy, but cannot source current, neither in a passive way (like capacitors and inductors) nor in an active way (like current sources).

Therefore, we propose here a hybrid information processing solution, which at the same time, will be functionally closer to the nervous system (Figure 2). Along with memristive synapses, we postulate the introduction of new types of elements that can source electric current in the form of pulses, either regular or chaotic/stochastic. Such elements should also respond to electrical signals applied in a way similar to living neurons, e.g. by changes in their pulsing frequency. A brief search over recent chemical literature brings candidates for this quest – electrochemical oscillatory processes.[27-28] Among the plethora of electrochemical oscillators those based on liquid metals: mercury, gallium and gallium-based alloys seem to be especially useful.[29-32]

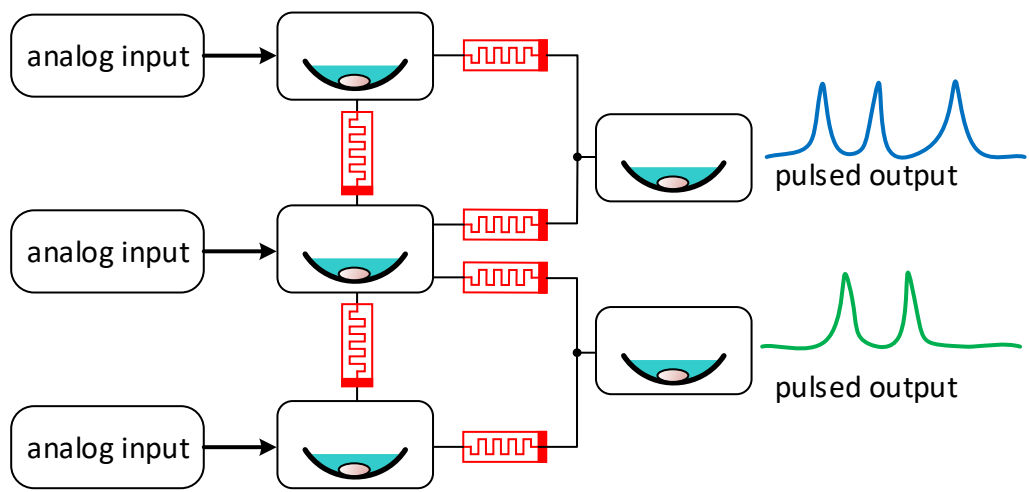

**Figure 2.** A tentative scheme of a memristively-coupled network of electrochemical oscillators. Depending on the state of the memristors (ON vs. OFF), various coupling modes will be present in the network, leading to a plethora of time-varying output responses forced by analog inputs.

The current perspective presents some ways of application of liquid metal-based artificial neurons and memristive synapses as two cornerstones of neuromimetic computing devices, for which we have coined the term *synthetic neural networks*. Liquid metal droplets are prone to both spontaneous and stimulated oscillations, which can be monitored by optical (change of shape and light reflectivity) and electrical (change in electrochemical potential, generation of galvanic current, etc.). Due to the dynamic properties of these oscillations, we coined the term *synthetic neurons* to describe electrochemical oscillators, which are stimuli-responsive, as they may show neuron-like behavior. Furthermore, it was demonstrated (in the case of mercury) that oscillating metal droplets can be synchronized, using resistive coupling, that regulates the degree of cooperation[33-34] and these synchronized systems behave similarly to neural systems (e.g., they show various catastrophic synchronization modes).[35]

Both systems seem to be electrically compatible: memristors usually operate with voltages ranging from hundreds of millivolts with currents ranging from nanoamperes to milliamperes, whereas electrochemical oscillators generate voltage amplitudes of hundreds of millivolts and can source at least microampere currents. The operating frequency of electrochemical oscillators reached hundreds of Hz at the maximum, so it is also well suited for semiconductor-based memristive devices.

The realization of the proposed hybrid devices and the synthetic neural networks requires three main building blocks: i) memristors and ii) electrochemical oscillators that operate in the specific time and potential scales, and iii) theoretical models to describe the operation of the system. Significant work was already done in the topics mentioned above, especially in the field of memristive devices and modeling of spiking neural networks. The main difficulty in the realization of the proposed concept will be the connection of all these pieces together. Attempts to create larger systems based on memristors or electrochemical oscillators have been already carried out (see chapter 3). The analysis of that partial result is crucial in the design of memristor-oscillator coupled systems. Below, we briefly summarize the basics of essential elements that connected together should allow creating the new synthetic neural networks based on the neuron-like hybrid device.

### 2.1 Memfractors, memristors and modeling

A memristor is an element providing memory – a crucial factor in the *in materio* computing paradigm – and synapse-like plasticity to the hybrid memristor-oscillator device. A *memristor,* as the fourth fundamental passive electrical circuit element, was postulated by Leon Chua as early as 1971.[36] For a long time it was considered only as a scientific curiosity, but in 2008 for the first time, a physical device was linked to the memristor postulated by Leon Chua. R. S. Williams and coworkers at the HP labs presented a device based on a thin layer of $TiO_2$ sandwiched between Pt contacts,[37] which showed the expected pinched hysteresis loop predicted in L. Chua's work. A memristor (memcapacitor, meminductor, or, more generally, a memristive element) is a passive, non-linear circuit element the properties (resistance, capacitance, or inductance) of which depend on the element's history and the total electric charge that has passed through the element in particular. For instance, in the case of an ideal, current-driven memristor, its constitutive equations can be written in a generic way as (1-3):[38]

$$U(t) = M[q(t)]I(t), \qquad (1)$$

$$q(t) = \int_{-\infty}^{t} d\tau \, I(\tau), \qquad (2)$$

$$M[q(t)] = R_{OFF}\left(1 - \frac{R_{ON}}{\beta}q(t)\right). \qquad (3)$$

Here, $U(t)$ is the voltage across the element, $I(t)$ is the current, $M$ is the *memristance* (charge-dependent resistance), $q(t)$ is the electric charge that flew through the device up to time $t$, and $R_{ON}$ and $R_{OFF}$ are resistances in the so-called low-resistive and high-resistive conducting states, respectively. In this way, a memristor can remember its history since its resistance is a function of the total charge that passed through the device. More importantly, if the current is switched off, the memristance value remains unchanged. The memristive properties displayed by various solid-state devices can be a consequence of several dynamical physical and chemical processes, including the formation of conductive filaments,[39-41] the migration of ions and/or dopants,[42] or the modulation of Schottky barrier heights.[43-44] Hence, physically meaningful modeling of a memristor needs to take into account the specific underlying physical mechanism(s).

The latter of the cases mentioned above – Schottky barrier modulation – is especially relevant due to the simplicity of the Schottky junction fabrication; however, it is the rarest case of all memristive devices reported so far.[45-47] The fabrication of a Schottky junction is a complex problem involving interactions between contacted materials at the atomic level.[48-49] Therefore, previously proposed, simplistic models taking into account only macroscopic parameters (e.g., the work function[48-49]) cannot yield reliable predictions of Schottky barrier parameters. For instance, the formation of metal-induced gap states (MIGS) within the bandgap can sensitively alter the local electronic structure at the interface. Thus, depending on the charge neutrality level, MIGS can exhibit both donor or acceptor-like properties.[50-52] Mem-devices, based on a Schottky junction, have more complex characteristics than ideal memristors (Figure 3). The main difference is usually observed as the asymmetry of the hysteresis loop. These devices display synaptic plasticity, but to the best of our knowledge, their dynamic is different from currently implemented devices in large neuromorphic systems.

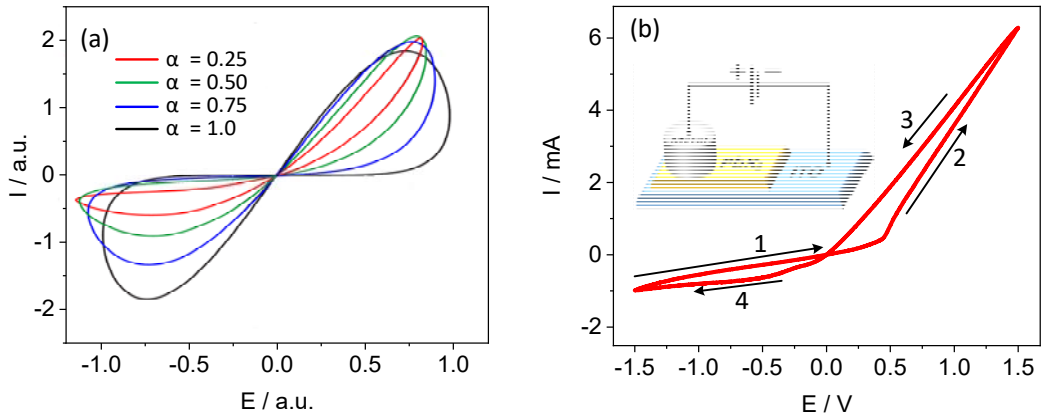

**Figure 3**. Pinched hysteresis loops for a charge controlled memristor for various values of the fractional-order $\alpha$ and a constant $\beta = 1$ value (cf. Eq. 6), adapted from Ref. [53] (a). The case $\alpha = 0.25$ (red curve) has a much closer similarity to the experimentally observed Schottky barrier in lead halide memristors (b). Adapted from Ref. [54].

Despite intensive experimental and theoretical works on memristive devices, they are still far from being fully understood and characterized, mostly due to the variability in the physical control parameters as well as in the underlying microscopic mechanisms leading to a display of memory effects in the devices. The most striking aspect is that the simple analytical description given by Eq. 1-3, which originates from dopant diffusion mechanisms, as well as other analytical and numerical models of memristors reported so far, cannot accurately predict the I/V characteristics and features of real memristive devices.

This has been the motivation to attempt a further theoretical generalization of memristive elements, which promises to achieve a closer connection with real experimental setups. This approach, presented by M. Abdelouahab, R. Lozi, and L. Chua, assumes non-integer order differential operators,[55-56] i.e., instead of using first-order ordinary differential equations it exploits the concept of fractional derivatives and fractional integrals to gain flexibility in the description of memory effects. Without entering into mathematical details, a fractional derivative (in the so-called Riemann-Liouville sense) is defined through the expression (4):[57]

$$_aD_t^\alpha f(t) = \frac{1}{\Gamma(n-\alpha)} \frac{d^n}{dt^n} \int_a^t \frac{f(\tau)}{(t-\tau)^{1+\alpha-n}} d\tau, \tag{4}$$

with $\alpha$ being the fractional order of the operator, $n = \text{int}[\alpha]$, and $\Gamma(.)$ is the Gamma function given by (5):

$$\Gamma(z) = \int_0^\infty x^{z-1} e^{-x} dx \tag{5}$$

The range of temporal correlations can be tuned by the power-law integral kernel $(t-\tau)^{-1-\alpha+n}$.

Based on this approach, L. Chua[56] and others[53,58] extended the definition of memristors and other similar mem-devices, introducing the concept of memfractance $F_M^{\alpha\beta}(t), (\alpha, \beta \in \mathbb{R})$ and a generalized formulation of Ohm's law in terms of fractional derivatives (6):

$$v(t) = D_t^{1-\alpha} \left( F_M^{\alpha\beta}(t) D_t^{\beta-1} i(t) \right) \tag{6}$$

Depending on the values of $\alpha, \beta$ the standard capacitance, resistance, and inductance can be recovered as well as intermediate circuit elements displaying combinations of the previous basic elements. Using fractional derivatives to describe the memristor dynamics (thus defining a fractional-order memristor or memfractor) makes the theory not only mathematically elegant, but it also provides much better control over the shape of the pinched hysteresis loop (see Figure 3), i.e., it allows to achieve a more accurate description of memristive behavior in realistic devices.[53-54]

We remark that fractional calculus has also been extensively applied to describe anomalous diffusion processes in disordered media.[59-61] as well as to other natural phenomena displaying power-law scaling in their temporal behavior, see, e.g., the reviews by Chen et al.[62] and B. J. West.[63-64] Fractional calculus is thus conceptually consistent with charge carrier transport in materials with randomly distributed charge traps, where power-law behavior may emerge in a natural way. We notice, however, that directly applying eq. (4) to model memfractor behavior in specific materials, seems rather difficult due to its very generic character, so that memfractive behavior has only been described in a very phenomenological way using, e.g., simple series expansions in the powers of the electric charge, see Refs.[56, 65] for typical examples.

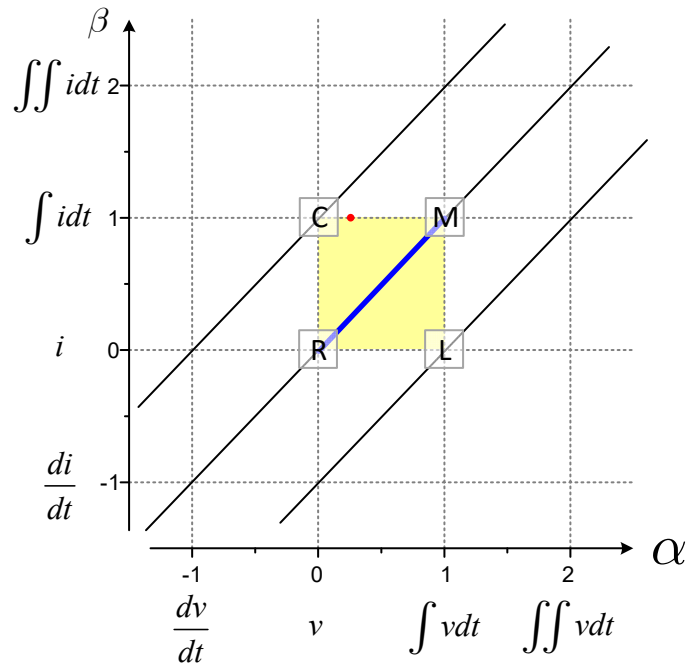

**Figure 4**. A fragment of a periodic table of memristive circuit elements defined on the space of the order of differential operators. R stands for a resistor, C for a capacitor, L for inductor, and M for memristor. The yellow area indicates the devices of interest discussed in this paper. The red dot indicates the position of a ½-order memfractor from Figure 3, whereas the blue line indicates the simplest memfractive elements that cannot accumulate energy; however, they still provide memory features significant for signal and information processing.

One of the possibilities of formal realization of memfractors involves networks of first-order memristors (cf. Eqs. 1-3 and Figure 4), with a memfractance order $\alpha$ depending on the network topology. In an example shown in Figure 5, a ½-order memfractor is built from first-order memfractors (i.e. $(\alpha, \beta) = (1, 1)$ memristors).[58, 66-67] More practically, an approach used for other types of fractional circuit elements can be applied with modifications, e.g., capacitor-like devices containing conducting particles dispersed in a dielectric matrix or so-called constant phase elements realized in electrolyte-based devices.[68]

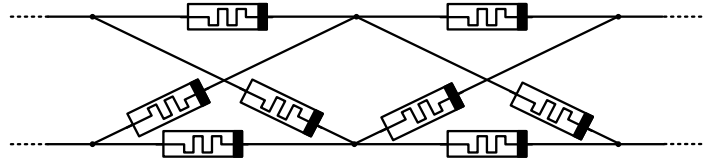

**Figure 5**. An equivalent circuit comprising of ideal memristors with a property of ½-order memfractance. Adapted from Ref. [58].

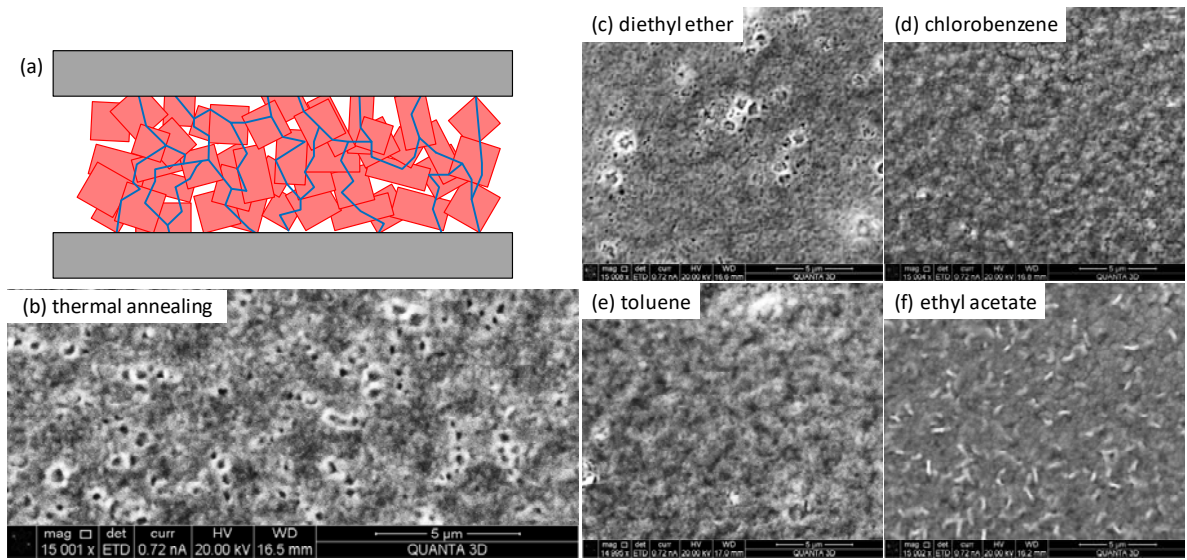

**Figure 6**. A cartoon illustrating the structure of a non-ideal memristor based on a polycrystalline material. Blue lines indicate some selected percolation pathways (a). SEM images on methylammonium triiodoplumbate thin layers on ITO glass obtained via thermal annealing (b) or in the presence of four different antisolvents (c-f). SEM images courtesy Tomasz Mazur and Piotr Zawal.

Therefore, we expect that polycrystalline, highly disordered semiconducting materials with a complex conductivity mechanism or appropriate arrangement of trap states materials may display memfractive properties. A specific example of a material class where the concept of memfractance may find an immediate embodiment are polycrystalline and disordered structures based on oxide or perovskite-type semiconductors or other microcrystalline materials (Figure 6). In such materials, each crystallite (especially if in contact with the metallic electrode) may work as an individual memristor. Therefore, the bulk of the materials may be considered as a complex, percolated network of memristive elements with complex charge transport properties,[69-75] in which the memfractive properties should emerge. The observed discrepancies between experimentally recorded and simulated *I-V* curves (cf. Figure 3)[76] may be a hint that the devices based on disordered structures are much closer to memristive networks (i.e., memfractors) than to ideal memristors.

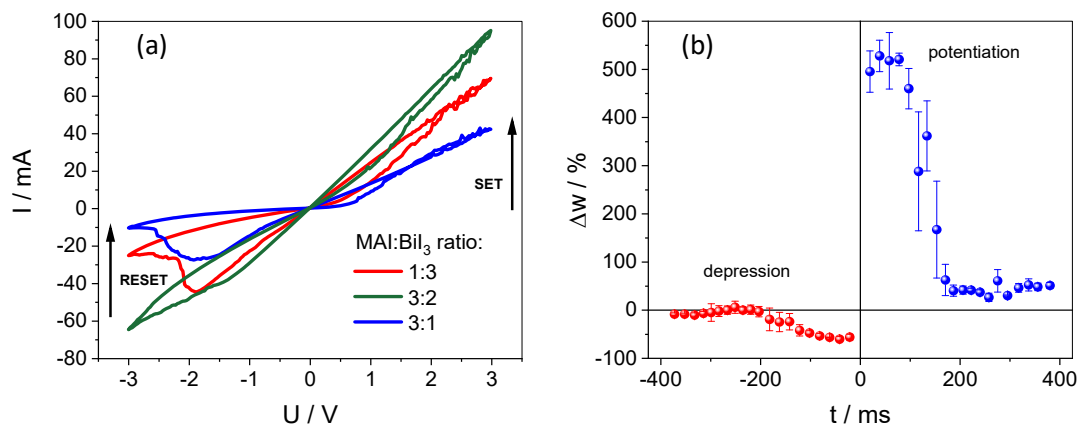

It was already found that thin layer memristive devices with significantly asymmetric I/V characteristics, similar to theoretically obtained characteristics of memfractive elements (cf. Figure 3), exhibit synaptic plasticity and follow typical Hebbian-like learning pattern, but their learning profiles are highly asymmetrical (Figure 7). In the studied case the depression mode of synaptic weight change was by ca. one order of magnitude smaller than the corresponding synaptic weight increase in the potentiation mode.[77]

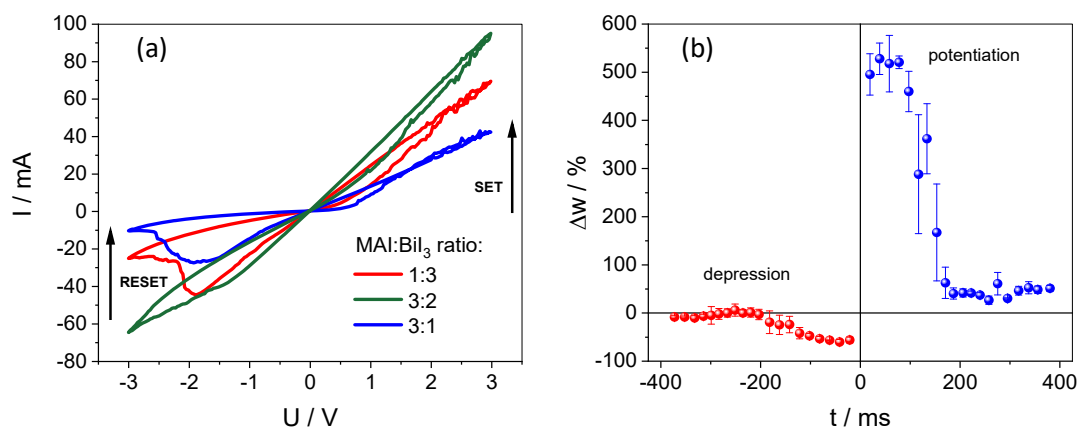

**Figure 7**. Asymmetrical pinched hysteresis loops recorded for methylammonium iodobismuthate layers with various methylammonium iodide (MAI) and bismuth iodide ratios (a) and the spike-timing-dependent plasticity for the 3:1 material. Adapted from Ref. [77] with permission of The Royal Society of Chemistry.

Coming back to the modeling of memristive devices, most of the work carried out so far has primarily focused on representing memristors by general phenomenological models which reflect the electrical response of a memristive device regardless of its physical structure or material composition.[78-81] Atomistic modeling has not been used very often and has mostly focused on the broad class of resistance switching related to vacancy migration in solid-state memristors, see, e.g., Ref. [82] for an overview associated with Valence Change Materials (VCM). Still, they have contributed to shed light on some microscopic, materials-related aspects of memristive behavior. Thus, Williams et al. used Molecular Dynamics (MD) simulations to investigate systems with oxygen vacancies driven by an external voltage.[83] The obtained results showed a competition between short- and long-range interactions among the oxygen vacancies, leading short-range ordering and mimicking memristor polarity inversion. Similar MD simulations highlighting other aspects of memristive behavior and ion dynamics can be found in recent literature.[84-85] Access to longer time scales relevant for vacancy dynamics can be reached via kinetic Monte-Carlo (kMC) simulations. A kMC approach was used to investigate structural configurations of a $TiO_2$ memristor, with a focus on oxygen vacancy migration under an externally applied voltage.[86] The authors found that vacancy hopping-induced localized electric fields seem to play an important role in the structural evolution of the filaments. Interestingly, using Density-Functional Theory (DFT), they found that filament induced gap states can emerge, leading to an insulator-metal transition during the filament formation process. Further DFT-based investigations of structural and thermodynamic features of vacancies in various memristive systems were presented recently.[87-90]

Phenomenological models, on the other side, are clearly sufficient for circuit-level simulations but are limited to specific switching mechanism(s) of the memristor and include various physical parameters that describe the active material characteristics in the device, thus reflecting potentially different switching mechanisms. We believe that a thorough understanding of atomic-scale processes leading to memristive behavior is fundamental in order to optimize material properties and help in the design of memory-based devices. Moreover, bridging atomistic information, encoding material properties, with phenomenological memristor (or more general memfractor) models represents a formidable challenge to modeling approaches, which deserves to be addressed in order to open the road for rational materials design.

A fascinating aspect of memristive elements, going beyond their relevance in circuit design, is that they can be considered as an electronic equivalent of a synapse. With that in mind, various applications for signal processing and classification have been developed.[91-93] Memfractive devices should share similar features; moreover, they should exhibit much better flexibility and memory features, as the character of the hysteresis loop should be tunable by structural parameters of the material (cf. Figure 2).

## 2.2 Electrochemical spiking neurons

The second part of the hybrid synthetic neuron is an electrochemical oscillator that connected to the memristor will modulate the resistive state of the former with periodic spikes. The famous Hollywood movie "Terminator 2: Judgment day"[94] presents a science-fiction vision of liquid metal-based technology, soft robotics, and artificial intelligence. Due to the unusual properties of room temperature liquid metals, they are the basis of emerging technology – liquid metal soft robotics.[29] This technology is based on mercury (the only room temperature liquid metallic element), gallium (melting point of 29.7°C), and gallium-based alloys (Ga-In eutectic, galinstan) with melting points in the range of 11-16°C. The main features of liquid metals that are explored in this context are high electrical conductivity, low viscosity (only twice the viscosity of water), and variable surface tension: high for metals with pure surface and decreasing dramatically in the presence of oxidized layers at the surface. Along with the electrocapillarity effect, this creates the possibility of shape and movement control of metallic droplets due to changes in the properties of the metal-electrolyte interface.[95-98] Whereas the mechanical properties of liquid metal interfaces are deeply explored, and new technologies based on this foundation have emerged, the information-related aspects of these phenomena have remained unexplored.

The relation between surface tension and the electric potential of the liquid metal surface was first described by Lippmann in 1873,[99] and the phenomenon is referenced as the electrocapillary effect. The "Mercury beating heart" – a classical chemical experiment is a beautiful and convincing demonstration of the electrocapillarity effect.[32] In this experiment, a mercury drop is placed in an acidic solution of an oxidizing agent (e.g., potassium chromate of permanganate) and touched with an iron nail. The mercury droplet starts to shrink and spread periodically – the "mercury heart" starts to beat. The same process (but with a different rate or character of oscillations) can be observed in acidic electrolytes with dissolved oxygen and with other reactive metals, either in direct contact with mercury or connected by tungsten wire. Two phases can be distinguished in the cycle: (i) oxidation in a moderately acidic medium result in the formation of mercury oxide islands (and subsequently a compact layer), which significantly reduces the surface tension of a mercury drop, that gradually spreads at the bottom of the vessel. (ii) At some point, it touches a piece of iron (or other active metal), and the surface oxide layer undergoes rapid reduction. The reduction is accompanied by a swift increase in surface tension, which makes the drop to shrink. These spontaneous oscillations repeat until the reagents (oxidant in the solution and the reducing metal) are consumed. The frequency of oscillations strongly depends on several factors including the volume of the metal drop, the composition of the electrolyte, type of active metal and its exposed surface, temperature, etc.[31-32]

Placing of a mercury drop between graphite electrodes and applying an external potential difference is another way to induce electrochemical-mechanical oscillations.[30] Oscillating metal drops generate electrical signals of amplitude up to ca. 1 V because they serve at the same time as galvanic cells. Very recently, electrochemically induced oscillations were observed in the case of molten gallium,[100-101] whereas galinstan and gallium-indium eutectic alloys should behave similarly.[102]

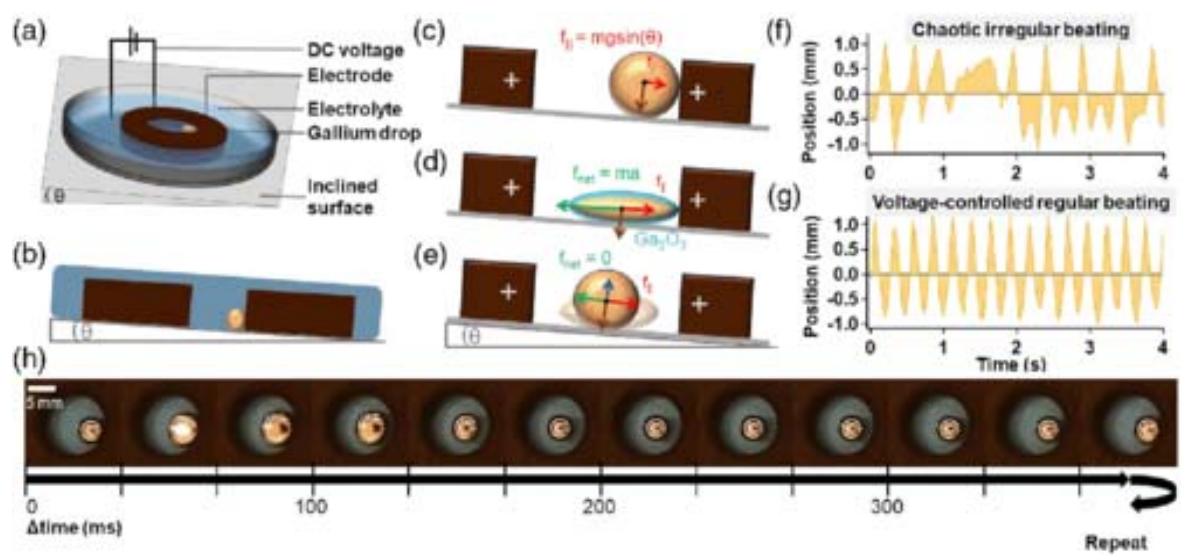

**Figure 8.** Schematic diagram of the experimental setup for the gallium oscillations (a). Cross-section of the inclined apparatus (b) and a force diagram of the gallium heartbeat sequence (c-e). Time course of chaotic motions occurring for certain drop sizes, or once the stabilizing voltage is removed (f). Regular periodic oscillations under a constant dc voltage observed for the 50 μL gallium drop (g). At the bottom, a series of photographs captured from the video shows the cycle of a heartbeat for the 50 μL drop (h). Reproduced from Ref. [100] with permission. Copyright American Physical Society, 2018.

Application of a constant DC bias to graphite ring electrode placed on inclined support induces strong and persistent oscillations of gallium droplets (Figure 8).[100] In the first step, the gallium droplet stays in contact with the electrode. If the electrode potential is higher than the standard electrochemical potential of gallium (related to reaction 7), the droplet gets oxidized at the surface:

$$2\,Ga + 3\,H_2O \rightarrow Ga_2O_3 + 6\,H^+ + 6e^- \qquad (7)$$

The gallium oxide layer thus formed acts as a surfactant, significantly decreasing the surface tension of liquid gallium. This results in the rapid spreading of the droplet at the inclined bottom of the cell and loss of electrical contact with the ring graphite electrode. Subsequently, the oxide layer is etched by an acidic or basic electrolyte. Removal of the oxide layer restores the high surface tension of molten gallium, which retains spherical shape, comes into contact with the electrode, and the cycle starts over. The frequency of gallium DC voltage-forced oscillations strongly depends on the volume of the droplet, applied voltage, and electrolyte composition (Figure 9). Depending on the applied forcing DC voltage and composition of the electrolyte two different types of behavior can be observed: irregular, chaotic oscillations (Figure 8f), and periodic oscillations (Figure 8g). The full map of various oscillatory modes is presented in Figure 9. The most important feature of the studied system from the point of view of neuromorphic information processing is the potential-induced controlled transition between chaotic and periodic modes. This allows us to define two dynamic states of the system. Furthermore, the periodic oscillations can be tuned over a certain frequency range. Such a feature makes the liquid metal electrochemical oscillator similar to a living neuron – the low-frequency background activity can be switched into periodic oscillation with a frequency depending on the intensity of the stimulus.

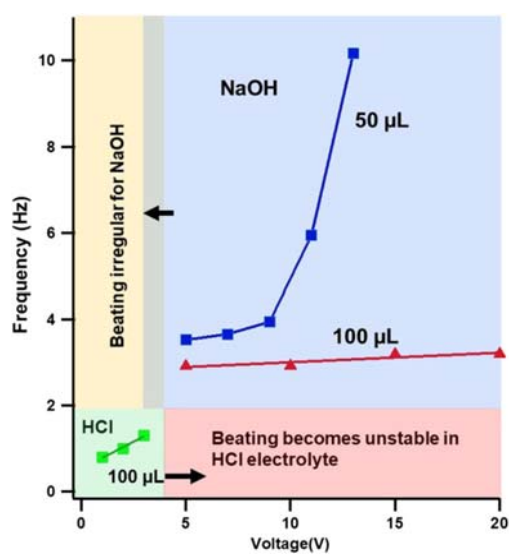

**Figure 9.** The beating frequency of a gallium droplet in NaOH and HCl electrolytes. Different threshold voltages are required to initiate beating in the acidic or basic electrolyte. Reproduced from Ref. [100] with permission. Copyright American Physical Society, 2018.

Upon application of an AC forcing signal, various oscillatory modes can be obtained, which are characterized by various shapes of oscillating metal drops and also accompanied by various electrical outputs.[103] Different topological modes of these oscillations usually generate specific signal patterns, which are best reflected in the complex Fourier spectra of the recorded electrochemical signals.[104] Moreover, electrically agitated liquid metal drops have already found application as unconventional circuit elements, but so far have not been coupled with the memristive device.[105-106]

### 2.3 Models of neuronal information processing

The human brain is definitely the most complex information processing system. Its unprecedented complexity is a result of approximately $10^{11}$ neurons connected into an extremely complex dynamic network with ca. $10^{14}$ synapses.[107] Each neuron collects information from numerous synapses, and depending on its own characteristics and the input (rate, amplitude, and time sequence of input spikes), it generates the output spike. Whereas the triggering of the action potential and propagation of signal along the axon is an electrical phenomenon, the information transfer and processing at synapses level relies on various chemical information carriers: neurotransmitters. Further complication originates from the fact that there are numerous types of neurons with various characteristics and there are numerous types of synapses and neurotransmitters.

One of the goals of the proposed hybrid device is to create a more complex synthetic neural network. This system, due to the presence of a spiking oscillator and memristor introducing plasticity, will be considered as a neuromorphic hybrid device. The modeling of the operation of the new type of network is one of three crucial components in the road to the application of such a system.

The development of the modeling of neural processes involves three phases, with increasing complexity and accuracy of modeling. The simplest networks are feedforward neural networks – systems mimicking the neural information processing based on a basic learning principle – modulation of synaptic weights between nodes and unidirectional information flow. The first, least advanced model is a perceptron – a first-generation neural network (Figure 10). It contains one input layer and one or more processing nodes. Each node calculates the weighted sum of inputs (8) and uses a simple threshold function (9) to generate the output, which has a binary character. Learning of the perceptron is achieved by updating all weights when known data are fed into the input in order to minimize the classification errors.

$$s = \sum_i w_i x_i \qquad (8)$$

$$f(s) = \begin{cases} 0, & \text{if } s > T \\ 1 & \text{otherwise} \end{cases} \qquad (9)$$

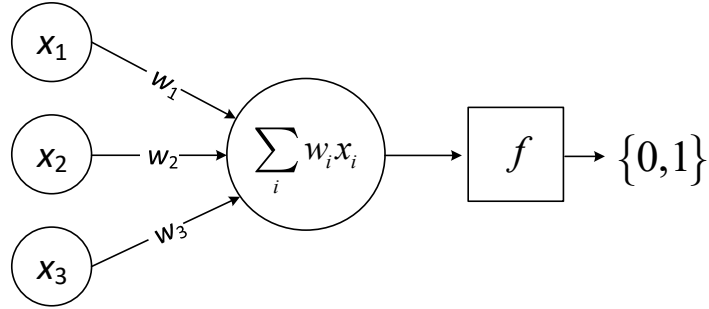

**Figure 10.** Schematic structure of a single unit perceptron network.

A more advanced approach (second generation neural networks) is based on the same geometry (connectivity) as the perceptron, but (i) it may contain more layers and (ii) it uses as an activation a logistic function (10) instead of a threshold function:

$$f(s) = \frac{1}{1+e^{-s}} \qquad (10)$$

This allows the generation of a continuum of output values and opens the possibility of multilayer architectures. Various activation functions can be used in different layers of the network; furthermore, a continuum of output states can be further processed using fuzzy logic approaches.[108]

The third, most advanced model of neural information processing is based on spiking neuron models (Figure 11).[107] Whereas the former models operate with information encoded in the amplitudes of inputs, the third-generation model is sensitive to the duration and time sequence of events, with the role of synaptic weights similar to those of the previous two generations.

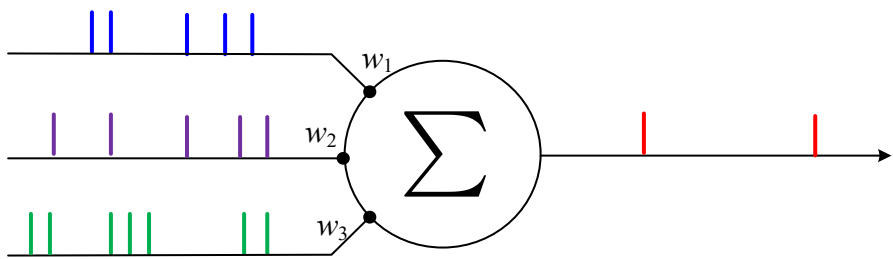

**Figure 11.** Schematic structure of single-unit spiking neural network.

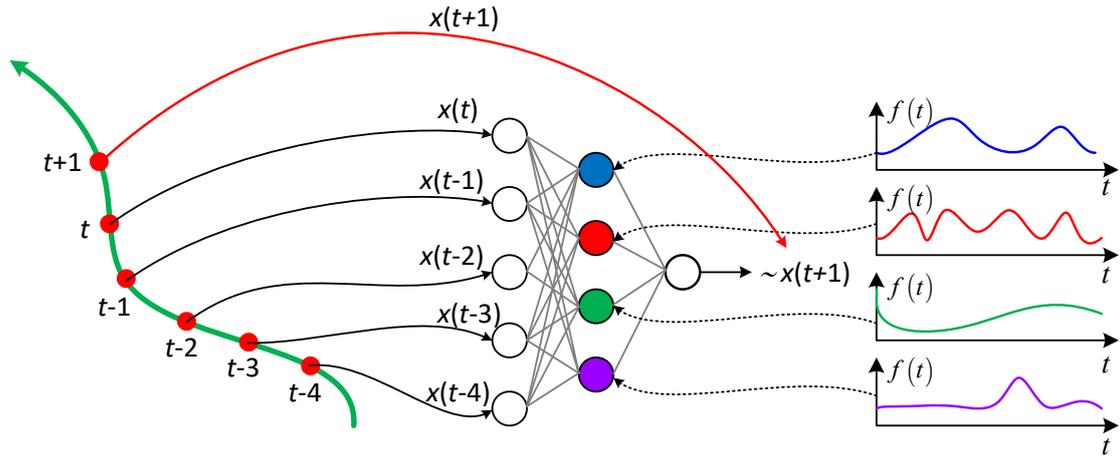

**Figure 12.** Schematic structure of a feed-forward neural network with nodes presenting time-dependent activation functions. Such a network can be used to predict the future dynamics of systems based on the system's history. Known trajectory at several points of time $\langle t-4, t \rangle$ serves a basis for approximate prediction of system dynamics at $t$+1. Adapted from Ref. [109].

Numerous mathematical models of spiking behavior can be applied, from the Hodgkin-Huxley,[110] leaky integrate-and-fire, Izhikevich,[111] Thorpe, and spike response models.[107] The common feature of all these models is their internal dynamics, leading to the generation of output spikes upon appropriate input conditions. These models also enable modeling of various coding approaches, as the rate (frequency) coding and pulse coding. In the first approach, the number of spikes within a defined time window encodes relevant information. In the second one, the exact time interval between subsequent events is the main carrier of information. Spiking neural networks also offer spike-timing dependent plasticity and spiking-rate dependent plasticity – two main components of natural learning mechanisms within the Hebbian learning paradigm.[112] These networks are not purely reactive, but also respond to the time sequence of events and therefore have a higher form of internal memory.

The highest level of performance can be achieved when the nodes of a network possess internal dynamics, e.g., when the activation function of the nodes varies in time. The simplest case of a dynamic neural network would be a perceptron-like system with a periodic change in the activation function of a single node in the network. In this case, the network automatically would become a phase- and frequency-sensitive filter operation in a similar way to a comb filter.[113] Any time-dependent periodic input signal will be processed by nodes with a periodically varying activation function. Therefore, if the input frequency will be commensurate with the eigenfrequency of the node, a positive interference will occur, and the two oscillations will be phase-matched. A network with numerous time-varying nodes would be capable of analysis of the time sequence of events and prediction of the future of input dynamics (Figure 12).[109] In this case, each node can extract one characteristic dynamic feature, and the combination of these features at several different time scales (defined by the rate of activation function variability) will be used for trajectory prediction.

Taking into account the quality and reliability of software spiking neural networks, it seems reasonable to look for experimental approaches towards them. While perceptrons and feed-forward networks can relatively easily be implemented with memristors (for details of memristors and their models see section 2.1 of this review), there are no physical realizations of artificial spiking neurons operating in a way similar to living neurons and based on hardware platforms different from standard CMOS technology. The photoelectrochemical neurons[114] reported so far have demonstrated their utility in certain pattern recognition tasks, including classification of hand-written digits.[115] Therefore there is an urgent need to develop spiking neuron hardware models based on unconventional substrates compatible with memristive synapses and other *in materio* computational systems.[116]

## 3. First steps toward larger systems

The idea of a hybrid memristor-electrochemical oscillator device is new, and up to our knowledge, it has not been realized so far at software or hardware level, neither as a single device nor in arrangements of individual components. Nonetheless, first attempts to create more complex systems with memristors or electrochemical oscillators have been made.

### *3.1 Memristive reservoirs*

Recently, a hardware implementation of complex memristive systems, especially reservoir systems, has attracted an increasing attention.[117-119] The main advantage of reservoir computers (as compared to, e.g., feedforward and recurrent neural networks) is the randomness of internal connections; it thus fits very well with the concept of *in materio* computing. Another advantage of reservoir systems is their flexibility: they allow effective processing of time-varying inputs and extract different features from large data sets by application of appropriate driving signals. Moreover, the application of delayed feedback may also convert simple neural networks or even single node synaptic/memristive devices into echo state machines, which are a specific kind of reservoir computer (Figure 13a).

Primarily, reservoir computing (RC) was developed to address the issue of the high computational cost of recurrent network training. Two forms of RC were independently developed, namely Echo State Network (ESN)[120] and Liquid State Machine (LSM)[121], where authors suggested the importance of multidimensional and rich dynamical state space. Later, the RC paradigm went beyond ESN and LSM, becoming a general model for information processing using non-autonomous dynamical systems with rich state space possessing a set of well-defined properties (*vide infra*).[122]

In general terms, a reservoir is a complex dynamical nonlinear system, the state of which at any point in time is a complex and unknown function of its history and has fading memory. The signal (the data to be analyzed) is supplied at the input and undergoes dynamic evolution at the reservoir nodes. The reservoir essentially maps the input into a point in a new space (created by the internal dynamics of the reservoir), thus performs a nonlinear transformation of the input.[123] Computation performed by the reservoir is represented as the projection of the trajectory of progressive states in its internal configuration space. In order to maximize the computing performance of the reservoir system, the user provides an additional input signal (also termed as "drive"), which needs to be tailored to maximize the distance between different regions of the configuration space into which the reservoir is driven. Fundamentally, we should have freedom in choosing the right drive, depending on the computational task we want to perform and the complexity of the reservoir layer itself. For a suitably complex reservoir layer, a relatively simple drive should suffice. [124]

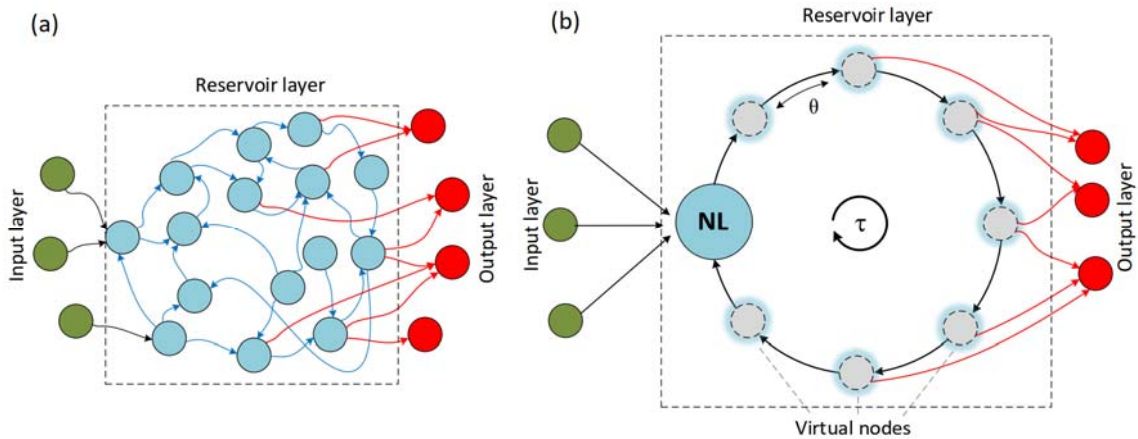

**Figure 13**. Schematic structure of a typical reservoir (a) of randomly, recursively and sparsely connected nodes, and the delayed feedback echo state machine (b) of circular topology. Green nodes represent different drives acting upon the reservoir. Red nodes represent different readout functions, depending on which property of the reservoir we are interested in probing. θ – denotes time between virtual neurons, i.e., the full time of delay line, τ – is the time needed for an efficient evolution of the system, NL – non-linear node. Adapted from Ref. [125]

The obtained different reservoir states can be then analyzed by the second part of the system, the readout layer (usually a simple perceptron or related system), which can be trained and is used to generate the final desired output. Training of the reservoir, in contrast to conventional neural networks, does not require any changes in the structure/synaptic weights inside the reservoir. The whole process involves only adjustments in the output layer and the drive. Therefore, reservoirs are considered as more efficient information processing devices than neutral networks, their training and operation is, however, more challenging.[126-128]

The dynamics of the internal reservoir state can be described with simple mathematical formulations.[129] In simplified terms, the reservoir evolves and is updated in time discreet steps $t$. It depends on two factors – its initial condition $x_{t-1}$ and input signal $u_t$, based on which current reservoir state $x_t$ can be obtained, giving (10):

$$x_t = \mathfrak{F}(x_{t-1}, u_t) \tag{10}$$

where $\mathfrak{F}$ is a function containing a description of the internal reservoir dynamics and its responsiveness to the input data. The readout layer can consist of any function $\psi$, depending on which reservoir parameter are we probing to obtain output information $y_t$, giving:

$$y_t = \psi(x_t) \tag{11}$$

The readout function $\psi$ probes only current states of reservoir $x_t$ (or just some fragments of it, cf. Figure 13) and – as it was stated above – this is the only part of the whole computing system that needs to be trained.

Reservoir computing systems only need to possess two very unrestrictive properties to achieve *universal computation power* for time-varying inputs: (i) the point-wise separation/generalization property (distant inputs signals should yield notable different states whereas close input signals should be mapped into similar reservoir states) and (ii) the approximation property for the readout function (readout function can map the current reservoir state to the desired current output with required accuracy).[93] We refer the interested reader to recent review papers[130-131] and specialized books.[132-134]

Important and closely connected features of reservoirs are their *fading memory* and *echo state property*.[135] Fading memory assures (through fading impact with the time of previous input on the current reservoir state) that operation of the reservoir does not turn chaotic, as it was shown that RC systems tend to be most efficient when operating at the edge of chaos.[136-138] The chaotic reservoir does not fulfill the generalization property so that even similar classes of the input signal are separated, and no classification can be performed. On the other hand, the echo state property - as the name implies - relies on the influence of the input signal on the dynamics of the reservoir, which is embedded in its future states. In other words, the current state of the reservoir configuration space depends on the echo of all previous input signals. In mathematical terms, the system of interest has an echo state property if there exists a function $\mathcal{E}$ formulated as (12):

$$x_t = \mathcal{E}(u_t, u_{t-1}, u_{t-2}, \ldots, u_{-\infty}) \tag{12}$$

which does not depend on the initial state of the reservoir, but only on the history of inputs. For an arbitrarily chosen systems, the existence of the $\mathcal{E}$ function cannot be guaranteed. Finite-time probing of the echo states on the basis of eq. 10 gives a recurrent recipe for the state of the reservoir (13):

$$x_t = \mathfrak{H}^{(h)}(u_t, u_{t-1}, u_{t-2}, \ldots, u_{t-h+1}, x_{t-h}) \tag{13}$$

For a given number of steps, it gives relatively simple recipes, examples for h = 2 and h = 3 are shown (14-15):

$$\mathfrak{H}^{(2)} = \mathfrak{F}(\mathfrak{F}(x_{t-2}, u_{t-1}), u_t) \tag{14}$$

$$\mathfrak{H}^{(3)} = \mathfrak{F}(\mathfrak{F}(\mathfrak{F}(x_{t-3}, u_{t-2}), u_{t-1}), u_t) \tag{15}$$

Whereas the existence of the $\mathcal{E}$ function cannot be guaranteed, the limit (16) can be safely taken for systems with an echo state property:

$$\mathcal{E} = \lim_{h \to \infty} \mathfrak{H}^{(h)} \tag{16}$$

The fading memory feature allows simplification for a finite interval of time (17):

$$x_t = \tilde{\mathcal{E}}^{(h)}(u_t, u_{t-1}, u_{t-2}, \ldots, u_{t-h+1}) \tag{17}$$

More detailed analysis of reservoir dynamics can be found in works of Jaeger and Konkoli.[120, 123, 139-144]

Memristors, due to their nonlinear characteristics and internal memory, may play the role of a node in neural networks and reservoir systems. The hardware implementation of a particular type of reservoir system, i.e., Single Node Echo State Machine (SNESM , Figure 13, right panel), was recently presented.[54] Figure 14a presents the experimental setup used to realize the idea of SNESM. The system, by definition, has only one physical node, and the feedback loop realized by incorporation of the delay line. The system also has a signal generator, a source-measure unit, and a recorder. The red arrows indicate the feedback loop created in the system. In general, both the signal and the state of the memristor change every time the signal passes through the device in each cycle.

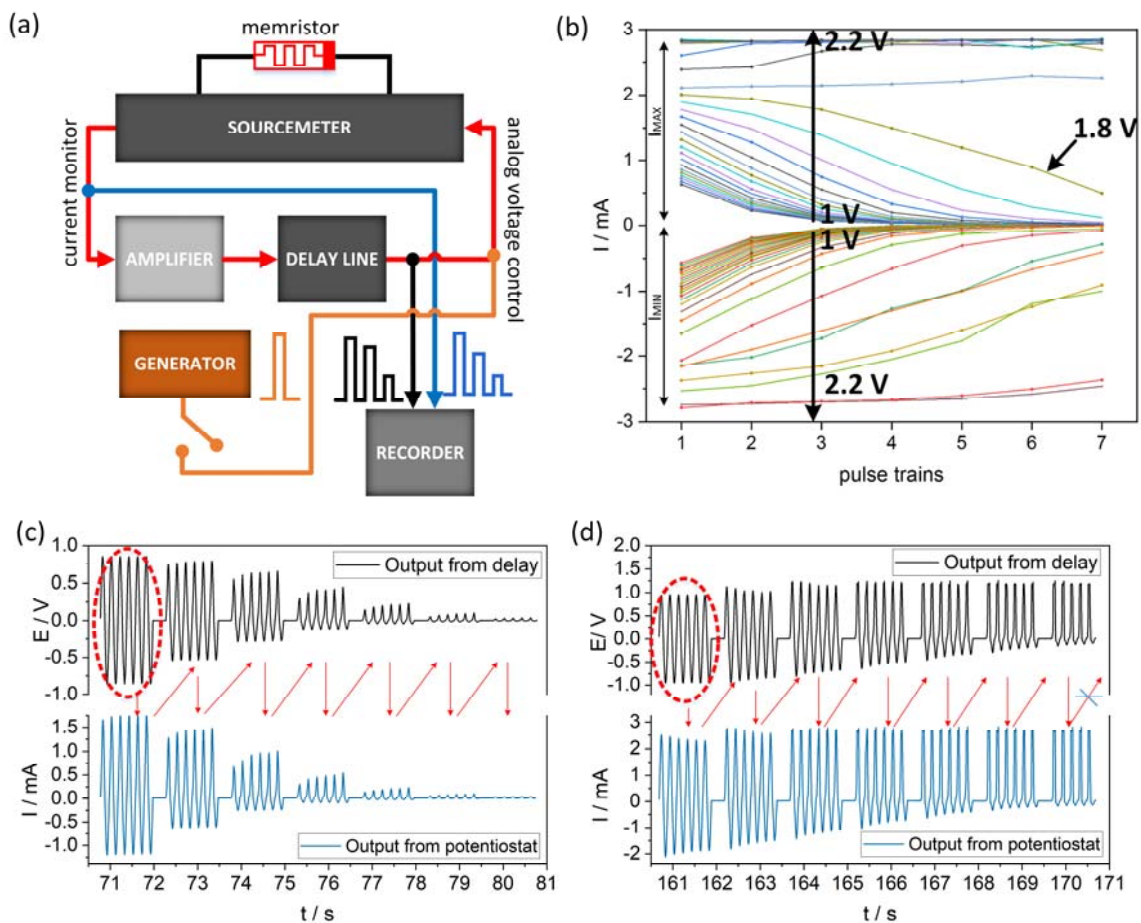

**Figure 14.** A scheme of the experimental system (a), the classification of the signal due to its amplitude (b), the two scenarios of signal circulating in the system (c), (d). Adapted from Ref. [54] with the permission of the American Chemical Society (ACS).

The presented system performs signal processing: the symmetrical signal passing through the node changes due to the asymmetrical *I/V* characteristics of the memristor. The signal is then delayed and returned to the same physical node several times. Figures 14 b-c show two possible scenarios to be observed during the experiment: the signal might circulate in the created loop several times and finally vanished. In the second scenario, the amplitude of the signal increases in each loop and can circulate in the system indefinitely. This feature of the presented Echo State Machine was used to perform a classification task. The set of sinusoidal signals of different amplitudes (in the range 1 to 2.2 Vpp, and 50 mV step) was introduced to the system, one by one. As a result, the amplitudes were divided into two subgroups: with the amplitude higher and with amplitude lower than a specific threshold value (here 1.85 Vpp, see Figure 14 b).

## 3.2 Coupled electrochemical oscillators

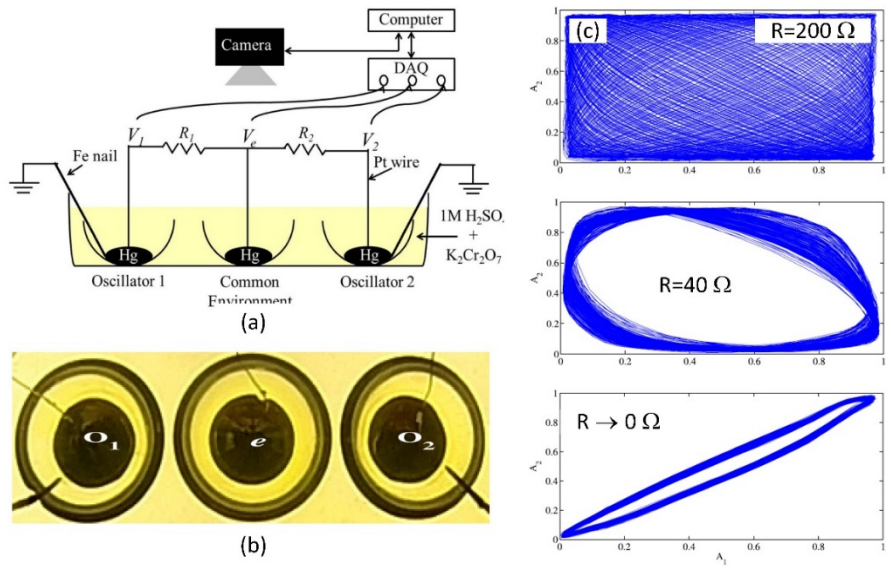

**Figure 15.** Schematic diagram of the experimental setup used to study the synchronization of individual electrochemical mercury beating heart system (a) and a top view of the experimental setup used to study the synchronization (b). State-space plots of the cross-sectional area of the Hg drop for different values of the coupling resistances. Reproduced from ref. [145] with permission. Copyright American Institute of Physics 2016.

The possibility of electrically forcing liquid metal oscillations along with their ability to generate alternating electrical signals of significant amplitude leads to a new research direction – studies in coupled electrochemical oscillators. There are numerous reports of coupling of mercury-based oscillators,[30-33, 35, 103, 145-150] but up to date, gallium-based systems were almost completely neglected. The behavior of gallium-based systems should be analogous to mercury-based systems, at least in terms of the dynamics and phenomenological description. It was demonstrated that two liquid metal electrochemical oscillators could be coupled via a resistive link (e.g., noble metal wires and resistor connecting two metal drops placed in the same[145] or in different[151] electrolyte baths). It was found in both cases that despite initial differences in drop volume (and hence differences in oscillating frequencies) the individual oscillators can couple in synchronous oscillation provided the coupling resistance is low (R→0), show partial synchronization with various phase shifts (R≈40 Ω) or oscillate independently (R≈200 Ω, Figure 15).[145]

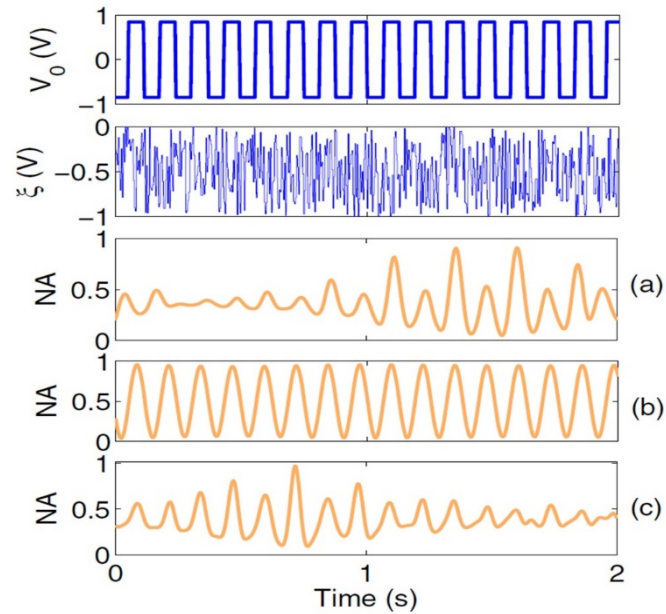

**Figure 16.** Demonstration of stochastic resonance in mercury beating heart system: $V_0$ is the subthreshold forcing signal, $\xi$ is the biased white noise signal. Lower curves show mercury drop oscillations (as a variation in normalized surface area of a drop) for various noise amplitudes: 0.2 V (a), 0.4 V (b), and 1.35 V (c). Reproduced from ref. [152] with permission. Copyright American Physical Society, 2019.

Application of operational amplifiers as coupling elements allows full control over the regulation of autonomous oscillatory activity of ensembles of oscillators,[146] as well as precise control of the oscillatory dynamics of individual oscillators.[153] Studies of larger assemblies of electrochemical oscillators with various connectivity indicated that these systems show various interesting phenomena. Depending on the frequency mismatch between individual oscillators and on their number, complex oscillatory patterns can be observed as a consequence of local coupling of oscillators with small frequency mismatch.[154-155] These partially synchronized networks may be regarded as functional analogs of polychronized oscillatory neural networks.[156-157] In this type of neural networks, due to the interplay between delays in signal transmission and plasticity, neurons of similar activity spontaneously self-organize into groups. The number of that groups changes dynamically and can even exceed the number of individual neurons, resulting in unprecedented computational performance. Electrochemical, liquid metal-based oscillators, coupled via dynamic (plastic) memristive links seem to be very well suited for computational purposes. This concept, however, has never been addressed, neither theoretically not experimentally, despite its experimental simplicity. Furthermore, such ensembles of coupled electrochemical oscillators show other features characteristic for neural networks (or other networks of coupled oscillators): explosive synchronization[147] and Kuramoto transitions.[148] Mercury-based weakly coupled oscillators are also prone to transition between periodic and aperiodic oscillations,[149-150] and show surprising sensitivity towards noise when maintained at specific oscillatory modes (e.g., triangular, cf. Refs. [103-104]).[152] This phenomenon can be explained in terms of stochastic resonance[158] and indicates the potential of liquid metal electrochemical oscillators as a sensing element (Figure 16). Furthermore, switching between various oscillatory modes can be potentially interesting from the IT technology point of view. This relevance is further supported by the specific features of weakly coupled oscillators – so called "cognitive modes", which have been already demonstrated for the Belousov-Zhabotinsky system.[159] The system of dynamically coupled oscillators can operate at the edge of chaos (the dynamic state, in which synchronization modes are highly unstable and can travel, in the phase space, between several basins of stability, or switch between various attractors).[138, 160-161]

On the basis of all the abovementioned features, especially the self-sustained oscillations, ability to couple with other oscillators, and responsiveness to external stimulation, electrochemical liquid metal-based oscillators seem to be ideal candidates for mimicking neural features in unconventional computing devices. In contrast to other unconventional computing systems, electrochemical oscillators should operate without external sources of power, as the oscillations are a consequence of electrochemical phenomena at the metal/electrolyte interface. The learning ability of such neurons will be, however, significantly limited. Therefore, they need to be coupled with synaptic devices with pronounces memory features. Whereas simple capacitive coupling can, in principle, mimic the short term memory,[162] coupling of electrochemical oscillators with memristive (or more precisely memfractive) elements will bring new computational features.

## 4. Towards autonomous robotics

Recent studies by S. Kasai and M. Aono have proven the utility of a simple, bioinspired computing device in the control of the autonomous robot.[163] This study was inspired by remarkable features of a primitive ameboid organism – slime mould *Physarum polycefalum*,[164-165] which is a model organism in unconventional computing studies.[166] The *Physarum* pseudopods are sensitive to light and other stimuli (e.g., food sources of chemical repellants) and change their geometry upon stimulation (Figure 17a).

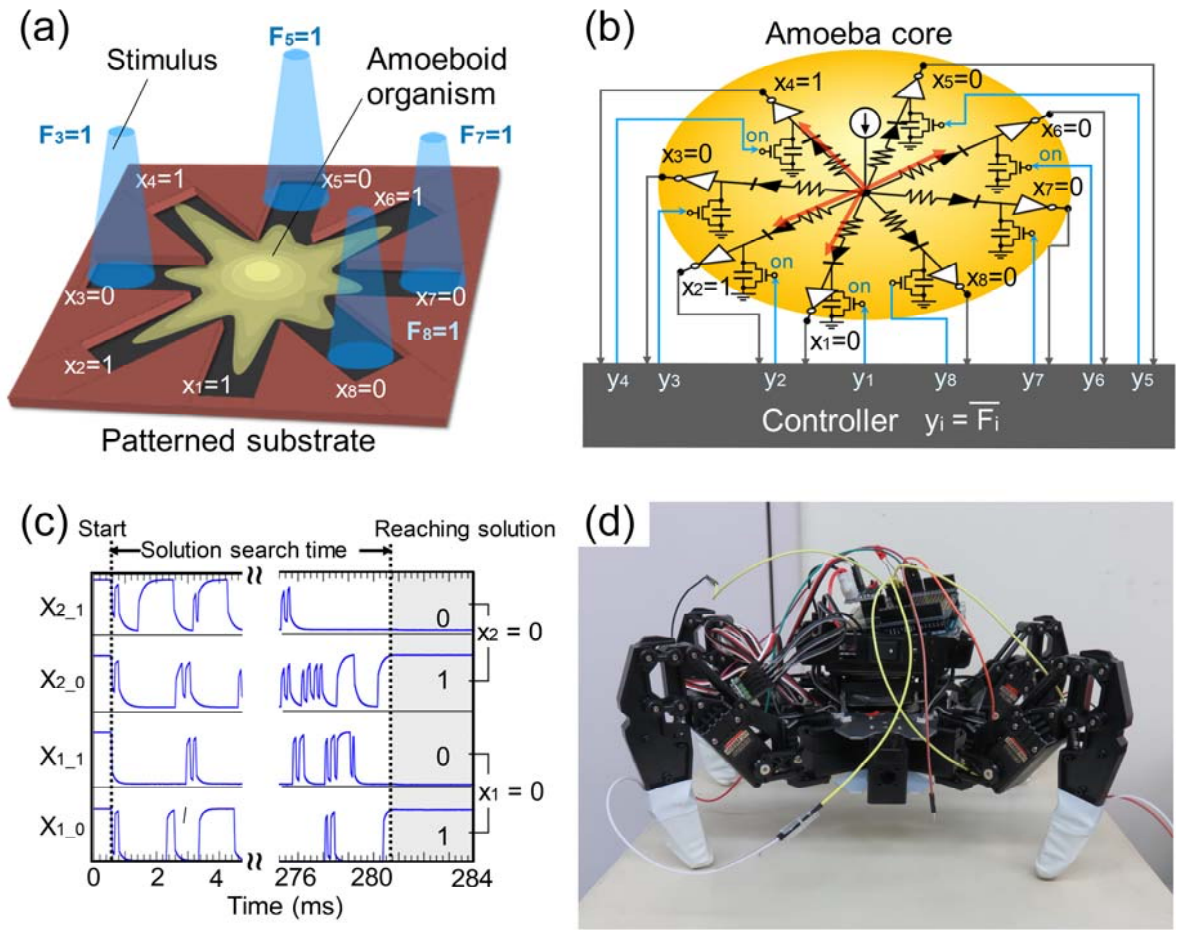

**Figure 17.** A basic concept of conversion of ameboid's response to stimulation to system control parameters (a), its implementation in a classical electronic circuit (b), the electrical signal recorded during the search for the solution of a leg movement (c) and an autonomous walking robot with an ameboid-inspired controller onboard (d). Adapted from Ref. [163] with permission. Original figures courtesy Prof. Seiya Kasai.

The electronic implementation of the ameboid-inspired robotic controller is based on a simple RC circuits with a FET transistor shorting the capacitor and an inverter serving as an output voltage source (Figure 17b). The controller maps the possible movements according to the Kirchhoff law according to current flow via the "pseudopods" of the circuit. Simple feedback from the touch sensors and application of the bounceback rule[165] provides signals (Figure 17c) that can control the movements of the robot (Figure 17d) successfully avoiding obstacles via solving the constraint satisfaction problem.

Here we propose that a similar control level can be achieved in an alternative way using unconventional computing devices based on electrochemical oscillators and memristive synapses coupled in a reservoir-like circuit without any digital electronic components involved in information processing. A set of two antiphase coupled liquid metal electrochemical oscillators should provide periodic signals triggering movements of legs on the left and right side of the tetrapod – with an appropriate delay between the front and rear legs. This should provide stimulation of the forwarding motion of the robot. Coupling of two oscillators should be realized with a memfractive element in an ON state. This will ensure the stable operation of synchronized oscillations and, thus, the steady speed and trajectory of the robot. Any obstacle detected by touch sensors modifies the rate of oscillations in one of the oscillators by changing the metal drop electrical potentials. Application of external potentials should, at the same time, switch memristor to higher resistivity thus decoupling oscillators responsible for left and right legs. Such a condition will switch an autonomous robot into a foraging mode, in which it will explore the neighborhood in a series of quasi-random moves. They will continue as long as the touch sensors will be disturbed by the obstacles. Finally, when the free path is discovered, the speed and trajectory of the robot should be stabilized by the coupling of both oscillators.

The numerical symilations[167-168] indicate the possibility of the application of a memristive element as a coupler between two oscillators. Moreover, more detailed studied clearly demonstrate the dependence of the degree of coupling on the state of the memristor.[169] Most probably, this control proposal should be extended to more memristive "pseudopods", like in Figure 17b. Constant phase shift, between oscillators, should be maintained by memristor-controlled all-pass filters. Details of this control paradigm will be developed in detail both on theoretical and experimental platforms. This concept is in line with the already explored field of liquid state machine-based robotic controllers, which show higher fault tolerance than another neuromimetic system. Despite the low speed of operation (50-250 Hz) of studied controllers, they have proven their utility in the learning of spatiotemporal behavioral sequences. They have been used a part of the autonomous robotic controller that was inspired by the structure and function of mushroom bodies – parts of insects' neural systems responsible for sensory functions and learning.[170] Application of memfractive devices within reservoir systems and coupled with artificial spiking neurons is in line with the common approach towards robot learning.[171]

## 5. Closing remarks

The analysis of apparently unrelated systems: liquid metal-based electrochemical oscillators and memfractors (or other memristive elements) reveals their utility in advanced information processing. Furthermore, it seems that they exhibit complementary functionalities.

Memfractors offer electrical nonlinearity, plasticity and memory features. They present synaptic properties and can be elements of responsive artificial neural networks. On the other hand, electrochemical oscillators provide dynamic oscillatory signals, the features of which (time profile, amplitude, frequency) are sensitive to the external stimuli (e.g., applied forcing potential, composition of the electrolyte). Their dynamics can be controlled at various levels and these oscillators can be easily coupled into larger systems, which exhibit criticality features similar to neural tissue.

Therefore we postulate a novel unconventional computing platform combining these two realms: artificial memfractive synapses and electrochemical spiking neurons. As a test bench, we suggest autonomous robotics, mainly due to the appropriate time scale of both types of devices.

## Acknowledgments

Authors acknowledge the financial support from the Polish National Science Centre within the MAESTRO (grant agreement No. UMO-2015/18/A/ST4/00058) and PRELUDIUM (grant agreement No. UMO-2015/19/N/ST5/00533) projects. DP has been partly supported by the EU Project POWR.03.02.00-00-I004/16.